\documentclass[%
print,
superscriptaddress,
 amsmath,amssymb,
aps,
floatfix,
twocolumn
]{revtex4-2}

\usepackage{graphicx}
\usepackage{dcolumn}
\usepackage{bm}
\usepackage{braket}
\usepackage{verbatim}

\usepackage[normalem]{ulem}

\begin{document}

\preprint{APS/123-QED}

\title{Synchronization of coupled Kuramoto oscillators\\ competing for resources}

\author{Keith A. Kroma-Wiley}
\address{
 Department of Physics \& Astronomy, University of Pennsylvania, Philadelphia, Pennsylvania 19104 USA
}

\author{Peter J. Mucha}
\address{
 Departments of Mathematics and Applied Physical Sciences, University of North Carolina, Chapel Hill, North Carolina 27599 USA
}

\author{Dani S. Bassett}
\address{ 
Departments of Physics \& Astronomy, Bioengineering, Electrical \& Systems Engineering, University of Pennsylvania, Philadelphia, Pennsylvania 19104 USA\\
}
\address{ 
Santa Fe Institute, Santa Fe, New Mexico 87501 USA\\
}



\begin{abstract}
Populations of oscillators are present throughout nature. Very often synchronization is observed in such populations if they are allowed to interact. A paradigmatic model for the study of such phenomena has been the Kuramoto model. However, considering real oscillations are rarely isochronous as a function of energy, it is natural to extend the model by allowing the natural frequencies to vary as a function of some dynamical resource supply. Beyond just accounting for a dynamical supply of resources, however, competition over a \emph{shared} resource supply is important in a variety of biological systems. In neuronal systems, for example, resource competition enables the study of neural activity via fMRI. It is reasonable to expect that this dynamical resource allocation should have consequences for the synchronization behavior of the brain. This paper presents a modified Kuramoto dynamics which includes additional dynamical terms that provide a relatively simple model of resource competition among populations of Kuramoto oscillators. We design a mutlilayer system which highlights the impact of the competition dynamics, and we show that in this designed system, correlations can arise between the synchronization states of two populations of oscillators which share no phase-coupling edges. These correlations are interesting in light of the often observed variance between functional and structural connectivity measures in real systems. The model presented here then suggests that some of the observed discrepancy may be explained by the way in which the brain dynamically allocates resources to different regions according to demand. If true, models such as this one provide a theoretical framework for analyzing the differences between structural and functional measures, and possibly implicate dynamical resource allocation as an integral part of the neural computation process.
\end{abstract}

\maketitle

\section{Introduction}
Oscillations are ubiquitous in nature. In many cases, populations of independent oscillators are weakly interacting with one another such that their intrinsic limit cycles are largely undeformed, but their positions along the limit cycles are pushed slightly forward or backward by their interactions with one another. In such situations, a common toy model for studying the observed phenomena is the Kuramoto model. The Kuramoto model describes a population of oscillators with intrinsic velocities $\omega_i$ interacting with one another through a sinusoidal coupling which tends to increase the instantaneous frequency of slow oscillators and decrease the instantaneous frequency of fast oscillators. Mathematically, the Kuramoto model is given by
\begin{equation}
    \dot\phi_i = \omega_i + \frac{K}{N}\sum_j A_{ij} \sin(\phi_j - \phi_i)\,,
    \label{plainKuramoto}
\end{equation}
coupling the dynamics of phase angle $\phi_i$ of oscillator $i$, with intrinsic angular velocity $\omega_i$, to its neighbors (encoded in adjacency matrix elements $A_{ij}$) by coupling constant $K$ in a system of $N$ oscillators.
The most interesting feature of the Kuramoto model is that the oscillators synchronize at a finite non-zero coupling strength $K_c$.

While the Kuramoto model has garnered significant interest and been used in a wide variety of applications, it is at its core a toy model. Few real systems follow Kuramoto dynamics precisely. Hence, many investigators have developed extensions to expand the dynamics in particular ways; such extensions include the inertial Kuramoto model \cite{Filatrella2008}, models in which natural frequencies and coupling strengths are time dependent \cite{Petkoski2012}, models in which the instantaneous velocities of oscillators feed back onto their natural frequencies \cite{Kroma-Wiley2021}, and models in which the local synchrony of an oscillator with its nearest neighbors in $A$ dynamically alters that oscillator's intrinsic frequency \cite{Nicosia2017}. In this vein, we previously considered a Kuramoto system that was modified to have a resource constraint, motivated by the non-isochronicity as a function of energy supply of most real oscillatory systems \cite{Kroma-Wiley2021}. Yet in many systems, for example neuronal systems \cite{Peters2004}, the oscillators are not only constrained by the availability of some resource, but they are also generally competing for the same limited supply of available resources. How resource competition affects synchronization dynamics is a natural question. 

Resource competition is undoubtedly an important feature of biological systems. We typically think of this competition as occurring between different animals competing for a limited supply of food or shelter. However, competition also exists within an organism. When exercising, the body must assess which muscle groups are most in need of energetic resources, and direct blood flow accordingly. So too in the brain, depending on the cognitive task at hand, the body dilates the vessels near the neurons which are most active, and so sends energetic resources to where they are most needed \cite{Raichle1976}. The redirection of resources necessarily means that some neural populations are deprived of the expected resource supply, simply because available resources are, at short time scales, fixed. There is a time lag of tens of seconds for heart rate and VO$_2$ to respond to increased demand, as shown in the case of exercise-induced demand \cite{Cooper1985}. Or consider the case of focal epilepsy, where a relatively small region of the brain becomes suddenly very active, and resources are directed there accordingly \cite{Kahane1999}. The impact of this resource redirection cannot be evaluated with a traditional Kuramoto model. 

The goal of the present work is to construct a model of synchrony that allows for resource competition between populations of oscillators. We study how competitive dynamics affect the observed behavior of the populations. In particular, we determine the critical coupling strengths necessary to induce synchrony in an unsynchronized population, and conversely, to eliminate synchrony in a synchronized population. We perform this investigation in the context of a specific oscillator network topology linked to a specific resource structure designed to highlight competitive effects. Further work could fruitfully extend the model to other systems with a range of distinct topologies and resource structures.

\section{Model}

\subsection{Resource-Constrained Kuramoto Model} The model we present here is an extension of that developed in Ref.~\cite{Kroma-Wiley2021}. In that prior study, we considered a system of coupled Kuramoto oscillators subject to resource consumption proportional to their oscillation rates. The resources were pulled diffusively from a fixed bath of the resource which was unique to each oscillator. That model can be written as follows:
\begin{align}
    \dot\phi_i &= \mu R_i + K\sum_j A_{ij} \sin(\phi_j - \phi_i)\,, \label{constraintPhi}\\
    \dot R_i &= D(B_i - R_i) + \beta \dot\phi_i\,. \label{constraintR}
\end{align}
In Eq.~(\ref{constraintPhi}) we have Kuramoto phase dynamics, with the natural frequencies of the oscillators all being proportional to their internal resource levels $R_i$. In Eq.~(\ref{constraintR}), we specify how these internal resource levels vary over time. In particular, each oscillator is connected to a bath of the resource at level $B_i$, and resources flow diffusively from the bath into the oscillator's internal resource supply. Further, resources are consumed (or produced, depending on the sign of $\beta$) in a manner proportional to the net frequency of the oscillator. 

To build the reader's intuition for this model (and how we extend it in the next section), we will briefly summarize the key behaviors and why they occur. First, we note that the resource constraint generates bistability in the synchronization dynamics when a positive feedback exists between oscillating faster and accumulating more resources (i.e., when $\mu$ and $\beta$ have the same sign). Intuitively, an oscillator has two equilibrium natural frequencies, depending on whether it oscillates at the group frequency or at its own natural frequency. When the feedback is positive, the equilibrium natural frequency is closer to the group frequency when the oscillator is entrained by the group, and further from the group frequency when it is free of the group. For an appropriate choice of coupling strength, then, both entrainment and independence can be stable behaviors for the oscillator. Hence a bistability in synchronization state arises for an individual oscillator, leading to multistability of the synchronization state of the total population. 

Alternately, the resource constraints can generate oscillatory dynamics when a negative feedback exists between oscillating faster and accumulating more resources (i.e., when $\mu$ and $\beta$ have opposite signs). In this case, the natural frequency of the oscillator moves away from the group frequency when the oscillator is entrained to the group, and toward the group frequency when the oscillator is free of the group. Thus, for appropriate choices of the coupling strength, neither entrainment state is stable for an individual oscillator. When the coupling strength is in this intermediate regime, the resource levels---and therefore the natural frequencies---tend to lie at the threshold natural frequency which can be overcome by the Kuramoto coupling term, due to the continual transitions in and out of entrainment by individual oscillators. Since the threshold natural frequency exists symmetrically above and below the mean frequency, the natural frequency distribution becomes bimodal. It has been shown in Ref.~\cite{Pazo2009} that a Kuramoto system with a bimodal natural frequency distribution undergoes a subcritical bifurcation from asynchrony to synchrony, which means that the synchronization state is bistable. This bistability creates an inertia to changing the global synchronization state, since the synchronization state will only change once the parameters (in this case, the natural frequencies) have drifted so that one of the two stable states vanishes. This inertia to global state changes tends to gradually induce a pacemaking mechanism in the state changes of individual oscillators, and so a global cycle arises. For more in-depth discussion of this resource-constrained model, we refer to Ref.~\cite{Kroma-Wiley2021}.

\subsection{Resource-Competition Kuramoto Model}

In contrast to the resource-constrained Kuramoto model just discussed, here we consider the more physically sensible situation in which the oscillators are competing for a shared influx of resources. In particular, we allow the bath levels to vary over time, always keeping the total bath level of the system fixed. This will serve to keep a constant influx of resources, but allow the system to distribute that influx differently among the oscillators. A dynamics that captures time variation while conserving total supply is the dynamics of a random walk. Since we also want resources to travel to where they are most needed, we will bias this random walk. We could bias the walk toward oscillators with the lowest resource levels $R_i$, which would drive a standard resource diffusion process. However, taking neurons as the system of interest, prior work has shown that blood vessels dilate in response to neural activity through a number of pathways, many of which are direct couplings between neurons or astrocytes and the vessels themselves \cite{Filosa2007}. This direct neuron-to-vessel coupling motivates us to bias our random walk toward oscillators with the highest frequency $\dot\phi_i$, as these oscillators represent the most active neurons.

Mathematically, a continuous-time and continuous-quantity biased random walk is described by
\begin{equation}
    \dot B_i = p\sum_j C_{ij} (B_j b_i - B_i b_j)\,,
    \label{eqn4}
\end{equation}
where $B_i$ is the bath level of the $i$th resource, $p$ is a rate constant, $C_{ij}$ is 1 if the bath nodes are connected and 0 otherwise, and $b_i$ is the bias term, which is large when resources are biased toward node $i$. Since we are taking the bias to be toward the fastest oscillators, we take $b_i = \dot\phi_i$. Finally, since Eq.~(\ref{eqn4}) only determines how the bath levels travel between oscillators, we must also specify the total supply of bath levels in the system. That is, we must supplement Eq.~(\ref{eqn4}) with the additional statement that $\sum_j B_j = B_{\text{tot}}$. Thus, we can write 
\begin{align}
    &\dot\phi = \omega_i + \mu R_i + \frac{K}{N}\sum_j A_{ij} \sin(\phi_j - \phi_i)\,, \label{eqn5} \\
    &\dot R_i = D(B_i - R_i) + \beta \dot\phi_i\,, \\
    &\dot B_i = p\sum_j C_{ij} (B_j \dot\phi_i - B_i \dot\phi_j)\,, ~\mathrm{and} \\
    &\sum_j B_j = B_{\text{tot}}\, ,
\end{align}
as the final form of our resource-competition Kuramoto model.

Note that Eq.~(\ref{eqn5}) differs from Eq.~(\ref{plainKuramoto}) by the addition of $\omega_i$, an oscillator-specific shift to the resource-dependent natural frequency. The motivation for this addition is that the bias term in a biased random walk cannot be negative (see Supplement \cite{SM}). Specifically, each oscillator is given a sufficiently positive $\omega_i$ to ensure that its frequency is initially positive. If all frequencies are initially positive, then the frequencies will remain positive for all time (see Supplement). This additional constraint does not alter the possible resource-constrained dynamics of Eqs.~\ref{constraintPhi}--\ref{constraintR} because any $\omega_i$ added to $\dot\phi_i$ can be absorbed into a redefinition of $B_i$ in that model, and hence adding $\omega_i$ as we do in Eq.~(\ref{eqn4}) is equivalent to simply choosing a different distribution of the bath levels in the resource-constrained model. Accordingly, the dynamics of our prior model are fully conserved in this model, excepting the dynamical variation of the bath levels. 

As a final step in formalizing the model, we non-dimensionalize the set of equations above to reduce the number of effective parameters. This step does not affect the dynamical landscape, but does change the magnitude of directions in the dynamical phase space (see Supplement). We also make explicit here that the $\omega_i$ are distributed according to a Gaussian distribution with mean $\bar\omega$ and variance $\sigma$. In practice, we chose $\bar{\omega}\gg\sigma$ so that the probability of $\omega_i < 0$ is vanishingly small. We then obtain the following system of equations:
\begin{align}
    &\dot\phi = \tilde\omega + \tilde\sigma\omega'_i + \tilde\mu R'_i + \frac{\tilde{K}}{N}\sum_j A_{ij} \sin(\phi_j - \phi_i)\,, \\
    &\dot R'_i = B'_i - R'_i + \tilde\beta \dot\phi_i\,, \\
    &\dot B'_i = p\sum_j C_{ij} (B'_j \dot\phi_i - B'_i \dot\phi_j)\,, ~\mathrm{and} \\
    &\sum_j B'_j = 1\, ,
\end{align}
where the modified parameters are as follows in terms of the original parameters:
\begin{align*}
    &B_i = B_\text{tot}B'_i\,, \\
    &R_i = B_\text{tot}R'_i\,, \\
    &\tilde\omega = D^{-1}\bar\omega\,, \\
    &\tilde\sigma = D^{-1}\sigma\,, \\
    &\tilde\mu = D^{-1}B_\text{tot}\mu\,, \\
    &\tilde{K} = D^{-1}K\,, \\
    &\tilde\beta = B_\text{tot}^{-1}\beta\,, \\
    &\omega'_i \sim \text{Gaussian}(0,1)\,, ~\mathrm{and} \\
    & t' = D t \, .
\end{align*}

\section{Designed System}
\subsection{Multilayer System Setup}
\begin{figure*}
    \includegraphics[scale=0.3]{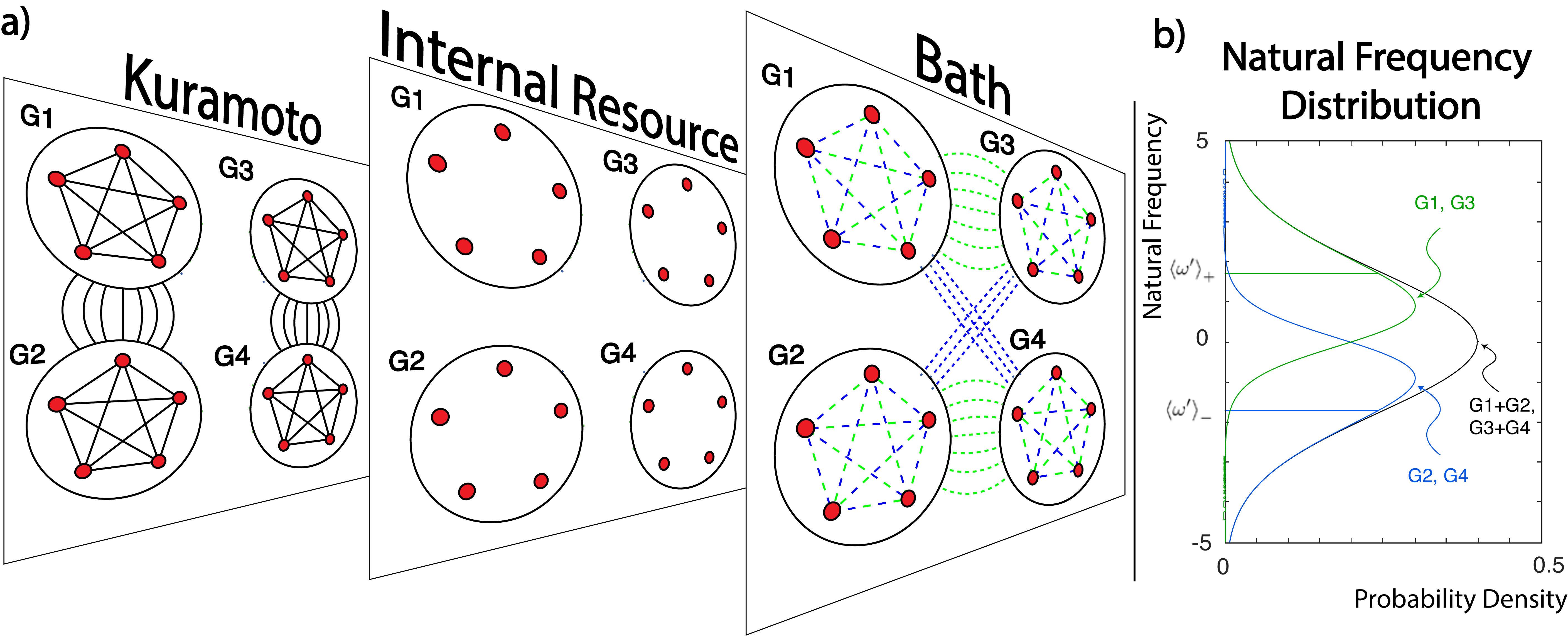}
    \caption{\textbf{System Overview.} (a) An overview of the system topology. The system can be visualized as a 3-layer network, with one layer for each set of dynamical systems. In the Kuramoto layer, all oscillators in groups G1 and G2 are connected together, and likewise for the oscillators in groups G3 and G4. The oscillators in G1 and G3 have $\omega'_i$ drawn from one distribution, while those in G2 and G4 are drawn from a different distribution, which is why they are divided from one another. In the internal resource layer, there is no connectivity, so no edges are drawn. Finally, in the bath layer, we consider two possibilities, indicated by either the green or blue dotted lines. In the blue case, which we call the high-low topology, all the oscillators in G1 and G4 are connected together, and likewise for G2 and G3. In the green case, which we call the high-high topology, all the oscillators in G1 and G3 are connected, and likewise for G2 and G4. (b) Here we show the probability distributions that the $\omega'_i$ are drawn from. We chose the distributions such that the overall distribution of $\omega'_i$ between G1 and G2 is Gaussian, but an oscillator in G1 is on average faster than an oscillator in G2 (likewise for G3 and G4).}
    \label{SystemOverview}
\end{figure*}
We are studying a system with a sizable number of parameters. In much of the parameter space, the competition dynamics in the bath layer are essentially static, amounting to a fixed, time-invariant distribution of the $B_i$ terms. In these settings the resource-constrained model would be sufficient to understand the dynamics. To illustrate the \emph{unique} behaviors of the resource-competition model, then, we must identify the small regions in the parameter space where competition dynamics are important. We do so in the context of a multilayer networked system that we design to have marked community structure. By understanding these relevant regions of the parameter space in this system, we gain tools and intuition for the study of more general systems.

The system we design contains four equally-sized groups of oscillators (see Fig.~\ref{SystemOverview}a): G1, G2, G3, and G4. Taking the groups in pairs, G1 and G2 will be all-to-all connected in Kuramoto layer $A$, meaning G1 and G2 are all-to-all connected internally, and every oscillator in G1 is connected to every oscillator in G2. Groups G3 and G4 will possess a connectivity in $A$ that is identical to that between G1 and G2. Further, we would like G1 to be all-to-all connected with either G3 or G4 in the bath layer, and for G2 to be all-to-all connected to the other of G3 and G4. When G1 is connected to G3 and G2 is connected to G4, we call that the high-high topology. When G1 is connected to G4 and G2 to G3, we call that the high-low topology.

Next, we will draw the frequencies $\omega'_i$ for each group according to the following probability distributions
\begin{align*}
    P(\omega')_{G1} = P(\omega')_{G3} &= (1+\tanh(\xi \omega'))\text{Gaus}(\omega';0,1) \\
    &\mathrm{and} \\
    P(\omega')_{G2} = P(\omega')_{G4} &= (1-\tanh(\xi \omega'))\text{Gaus}(\omega';0,1) .
\end{align*}
We see that in the $\xi\rightarrow\infty$ limit, this process amounts to drawing the $\omega'_i$ of G1 and G3 from the right half of a Gaussian distribution, and the $\omega'_i$ of G2 and G4 from the left half of a Gaussian distribution. For intermediate values of $\xi$, the division between the groups is less pronounced. Thus, if we consider G1 and G2 jointly, we observe that their overall distribution of $\omega'_i$ is Gaussian, and likewise for G3 and G4. Yet within the groups, one group will on average have the larger $\omega'_i$, and the other group will on average have the smaller $\omega'_i$. By connecting G1 with G3 in the bath layer, the higher $\omega'_i$ groups share a pool of bath levels, which is why we refer to this scenario as the high-high topology; by connecting G1 with G4, a higher $\omega'_i$ group shares a bath pool with a lower $\omega'_i$ group, which is why we call this scenario the high-low topology.

In addition to dividing the groups in this way and giving them distinct $\omega'$ distributions, we also give them different coupling strengths. In particular, we let the coupling strength in the all-to-all \{G1,G2\} network be $K_{12}$, and we let the coupling strength in the all-to-all \{G3,G4\} network be $K_{34}$. Recall that there are no Kuramoto connections in $A$ between \{G1,G2\} and \{G3,G4\}. Drawing this distinction will allow us to vary the coupling strengths separately to control the synchronization states of the two populations independently.

\subsection{Multilayer System Intuition}

To gain an intuition for how competition dynamics affect this system, suppose that both coupling constants $K_{12}$ and $K_{34}$ are initially zero. Without any coupling, both $G_1$ and $G_2$ will be completely unsynchronized, and all oscillators will obey their own natural frequencies $\tilde\omega + \tilde\sigma\omega'_i + \tilde\mu R'_i$. Now suppose that $K_{12}$ is increased until the \{G1,G2\} pair synchronizes, which it must for some large enough value of $K_{12}$. After synchronization, the G1 oscillators will be on average slower than their natural frequencies, while the G2 oscillators will be on average faster than their natural frequencies, since both groups have been pulled to the central group frequency. This behavior has immediate consequences in the bath layer of the multilayer network. When the G1 oscillators slow down, they reduce their own bias terms in the bath dynamics, thereby causing bath levels to generally fall in G1. Bath levels will likewise rise for G2. 

Importantly, these dynamics in G1 and G2 have immediate consequences for G3 and G4.
Specifically, the bath levels will rise for the group connected to G1 (i.e., G3 in the high-high topology scenario, or G4 in the high-low topology scenario). The increase in bath levels will cause the internal resource levels, and therefore the natural frequencies, to rise for one of these two groups. The bath levels for the group connected to G2 will fall. This process will have the effect of either pulling the natural frequencies closer together in the \{G3,G4\} pair (high-low topology), or pushing them further apart (high-high topology). In turn, this behavior will lower or raise, respectively, the critical coupling strength at which synchronization of the \{G3,G4\} pair occurs. By this cascade of effects, the synchronization state of the \{G1,G2\} pair will causally alter the synchronization state of the \{G3,G4\} pair, despite there being no Kuramoto coupling between them ($A_{ij} = 0$ when $i\in$ \{G1,G2\} and $j\in$ \{G3,G4\}). 

Thus, it may be impossible for both groups to simultaneously synchronize unless their coupling strengths $K_{12}$ and $K_{34}$ are raised above the critical coupling for either group to synchronize while leaving the other group unsynchronized. We expect such behavior for the high-high topology scenario, where synchronizing one group inhibits synchrony of the other group. In contrast, in the high-low topology scenario we expect synchronizing either \{G1,G2\} or \{G3,G4\} from the doubly-unsynchronized state (that is, neither group pair is synchronized) to require a greater coupling strength than would be necessary to \emph{maintain} the doubly-synchronized state (that is, both group pairs are synchronized). That is, in the high-low topology case we expect multistability, because it is easier to maintain the doubly-synchronized state than it is to reach it in the first place. Thus, if you raise the coupling until the doubly-synchronized state appears, then immediately turn it back down a bit, the doubly-synchronized state will persist, even though the coupling parameters are the same as they were just before the transition to the doubly-synchronized state.

\section{Results}

\begin{figure*}
    \includegraphics[width=7in]{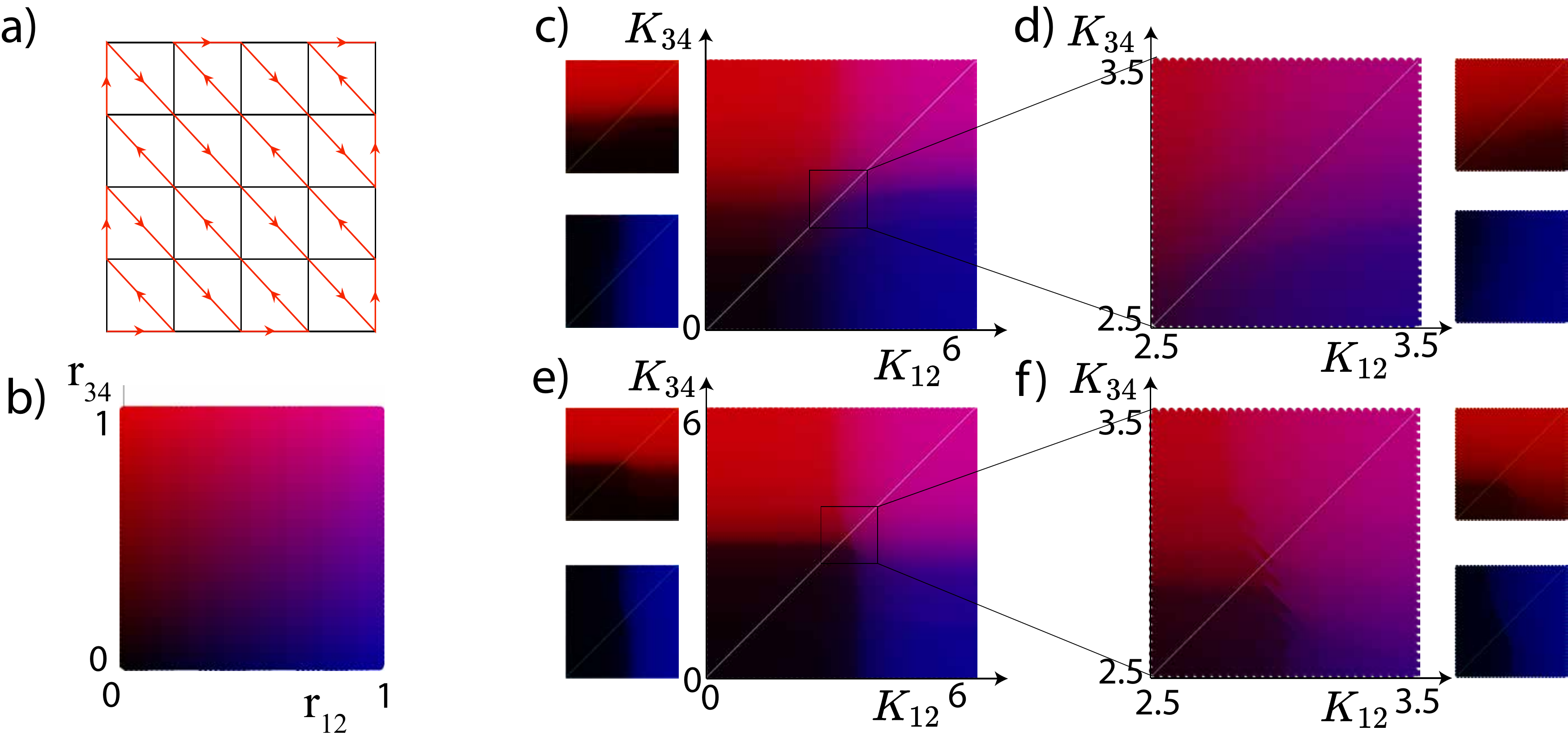}
    \caption{\textbf{System Behavior.} (a) Because we expect hysteresis to be significant in the model, we must choose a path through $\{K_{12},K_{34}\}$ space. Here we show the path we chose, which is essentially symmetric under group relabeling, as we expect the phenomena to be. In panel (b) we give a legend for interpreting the remaining plots. For each color in panels (c)--(f), the corresponding color may be found in panel (b), and the order parameter values may then be read off as the x and y coordinates of that color. In panels (c)--(f) we plot the synchronization state of the system. The main plot ties the amount of red to the order parameter of \{G3,G4\}, and the amount of blue to the order parameter of \{G1,G2\}. Thus, black indicates that both order parameters are low, red indicates high order in \{G3,G4\} but low order in \{G1,G2\}, blue indicates the reverse, and purple indicates high order in both group pairs. The smaller figures located at the sides of the larger figures isolate the red and blue channels separately, for the benefit of black-and-white printouts or colorblind readers. In panel (c), we plot the behavior for the high-high topology scenario, and observe the increased difficulty of maintaining the doubly-synchronized state compared to a singly-synchronized state, as evidenced by the shift of the critical coupling rightward and upward when one of the groups is synchronized compared to when neither group is synchronized. In panel (d), we zoom in on the critical region at the center of panel (c) and observe very little hysteresis, which would be indicated by adjacent lines differing significantly in color. Such color difference would indicate considerable difference in synchronization behavior depending on whether the region was approached from top-left or from bottom-right. In panel (e), we show system behavior in the high-low topology scenario and observe the relative ease of maintaining the doubly-synchronized state compared to either singly-synchronized state. Finally, in panel (f), we zoom in on the critical region in panel (e) and note that here we do observe significant hysteretic effects.}
    \label{Simulation}
\end{figure*}

\subsection{Structure of Simulations}

We first simulated the system with 1000 oscillators using the connectivity described above, with 250 oscillators in each of G1--G4. To isolate the multistability \emph{unique} to the resource-competition model, we studied the system in a parameter regime where dynamics induced by the simpler resource-constrained model were not bistable: $\tilde\mu >0$ and $\tilde\beta <0$. In particular, we chose the following parameter values for our simulation: $\tilde\omega = 0.6$, $\tilde\sigma = 0.04$, $\tilde\mu = 100$, $\tilde\beta = -0.03$, $p = 1\times 10^{-5}$.

To understand system dynamics, we simulated the system with adiabatic changes in $K_{12}$ and $K_{34}$ over time. The anticipated multistability of the system foregrounds the importance of the dynamics' history, which is why we increment the coupling strengths within the same system, rather than running the simulation anew for each pair \{$K_{12},K_{34}$\}. While the parameter values are the same for each instantiation, the initial conditions of the phases, resource levels, and bath levels are randomly chosen, as are the values of $\omega'_i$. To update $K_{12}$ and $K_{34}$ adiabatically, we must choose a path through $\{K_{12},K_{34}\}$ space. Considering the symmetry of $K_{12}$ and $K_{34}$, we chose to follow a symmetric path, which consisted of the diagonal up-left and down-right traversals (Fig.~\ref{Simulation}a). At each value of $K_{12}$ and $K_{34}$, we are interested in the order parameters of the \{G1,G2\} and \{G3,G4\} pairs, given by $r_{12}$ and $r_{34}$, respectively. The typical order parameter of a phase oscillator system---and the one we will use here---is given by $\frac{1}{n}\sum_j e^{i\phi_j}$ within each pair of groups, where $n$ is the number of oscillators in each pair.

\subsection{Observed Behavior}

Hence, we have two independent variables $K_{12}$ and $K_{34}$, and two dependent variables $r_{12}$ and $r_{34}$. We visualized this 4-dimensional space by placing $K_{12}$ and $K_{34}$ on the spatial axes of the usual $xy$-plane, and colored each point by encoding the value of $r_{12}$ in the amount of blue and encoding the value of $r_{34}$ in the amount of red (Fig.~\ref{Simulation}b). Intuitively, if both order parameters are small, then the point will be black; if both are large, then the point will be purple. When one order parameter is large and the other is small, the point will be either red or blue. We study two cases: in the first, the bath topology connects G1 to G3 and G2 to G4, which we call the high-high topology scenario; in the second, the bath topology connects G1 to G4 and G2 to G3, which we call the high-low topology scenario. As discussed previously, we expect simultaneous synchrony of both groups to be frustrated in the high-high topology case and made easier in the high-low topology case. Our results are consistent with our expectations. First, consider the high-high case shown in Fig.~\ref{Simulation}b. If we follow the phase transition of the \{G1,G2\} pair---indicated either by a transition from black to blue or from red to purple---we note that the phase transition moves to a higher coupling strength when the \{G3,G4\} pair is already synchronized than when the \{G3,G4\} pair is unsynchronized. This behavior is indicated by the black-to-red transition line occurring at a smaller value of $K_{12}$ than the blue-to-purple transition line. Second, consider the high-low topology case, where we observe the reverse phenomenon: the blue-to-purple transition occurs at smaller $K_{12}$ than the black-to-red transition (Fig.~\ref{Simulation}d). 

Now let us look at the zoomed in images on the right of the figure (Fig.~\ref{Simulation}c,e). In the high-high topology scenario here, we do not see complex behavior beyond that observed in prior work. In the high-low topology scenario, however, new features emerge. In particular, we observe clear hysteretic behavior. The hysteresis is most pronounced when the doubly-synchronized state is involved. In that case the purple of that state persists farther into the red and blue regions when we come \emph{from} the doubly-synchronized state than when we \emph{approach} the doubly-synchronized state. However, weaker hysteretic effects are present even in transitions between the doubly-unsynchronized and singly-synchronized states.

We can understand the doubly-synchronized hysteresis as a consequence of the cooperative synchrony in the high-low topology scenario. The hysteresis between the singly-synchronized and doubly-unsynchronized states, however, requires a distinct argument.
If we begin with both groups unsynchronized, and then one group synchronizes, the faster oscillators will move slower, and the slower oscillators will move faster, since all oscillators now oscillate at an intermediate frequency between fast and slow. Thus, in the bath dynamics, the fast oscillators will bias resources away from themselves after synchronization, while the slow oscillators will bias resources toward themselves. After synchronizing, the intrinsic frequencies of the oscillators will tend to move toward the mean due to resource redistribution, and this effect will tend to reinforce the synchronized state of the system. Hence, when the coupling strength is once again reduced, resources will now be distributed such that synchronization is easier, and so the coupling strength at which the downward transition occurs will be smaller than that at which the upward transition ocurred. This argument is independent of the high-high or high-low connection topology scenario, and so we would expect to see the effect in both cases. That said, although the reasoning applies to both cases, it does not mean that the magnitude of the effect is equivalent, and we will in fact find that the effect is much stronger in the high-low topology scenario. This explains why the effect is not very apparent in the high-high topology scenario shown in Fig.~\ref{Simulation}.

\section{Analysis}
Having established the qualitative behavior of this model, let us now try to quantify. Due to their length, most derivations will be relegated to the Supplement, but the results of interest will be presented here. 

We would like to understand the locations of the phase transitions in this model in $\{K_{12},K_{34}\}$ space. There are three limiting cases that we wish to consider analytically: the doubly-unsynchronized, the singly-synchronized, and the doubly-synchronized states. Based on the equilibrium intrinsic frequency distributions in these three cases, we will be able to understand roughly where the critical coupling strengths are in $\{K_{12},K_{34}\}$ space.

\subsection{Doubly-Synchronized}
We first consider the simplest case, when all oscillators are synchronized. In general, since our two group pairs are disconnected in the phase-coupling matrix $A$, we might expect that the two groups could have different synchronized group frequencies. However, as we show in the Supplement, no such equilibrium state exists. The only equilibrium possible is when both groups have the same average velocity. The reason is somewhat nuanced and depends on the $\tilde\omega$ term being non-zero. Essentially the bath resources are distributed in proportion to the phase velocities, but the velocities are not \emph{directly} proportional to the resources. Rather, they are linearly dependent on the resources with intercept $\tilde\omega$. It turns out that dependence forbids non-equal equilibrium solutions. Since the resource dynamics are bounded (there is a fixed total supply of bath levels), and there is only one fixed point, the group frequencies will either equilibrate to the fixed point, or oscillate around it. Proving the non-existence of oscillations would be challenging, but we do not observe them in the simulation.

Thus, when both group pairs are synchronized all oscillators turn with the same frequency independent of group membership. This behavior makes the resource distribution quite simple in functional form. Every oscillator will possess an equal supply of resources, which means the only variance in the intrinsic frequencies will be the variance $\tilde\sigma^2$ that we built into the model. Every other contribution to intrinsic frequencies is equivalent for all oscillators, and so only serves as a shift of the mean velocity, which does not impact synchronization behavior. Given that we know the variance, and we know that the distribution is Gaussian, we can use the usual formula for the critical coupling derived by Kuramoto \cite{Kuramoto1975}. Thus,
\begin{equation}
    K_c^\text{2sync} = \frac{\pi}{2g(0)} = \frac{\pi\sqrt{2\pi\tilde\sigma^2}}{2} \, ,
\end{equation}
where $g(0)$ is the probability density of oscillators having the average natural frequency $\omega_i + \mu R_i$. 

\subsection{Doubly-Unsynchronized}
The next case we will consider is when both populations are unsynchronized. As before, the complete derivation can be found in the Supplement. In the absence of coupling, the equilibrium state can be found fairly straightforwardly by setting $K_{12} = K_{34} = 0$, setting the time derivatives to zero, and then solving the resulting system of equations. In this case, the behavior depends on whether we have a high-high or high-low topology scenario in the bath layer. In the high-high topology scenario, the distribution of resources is no longer Gaussian within a group, as the frequency of an oscillator picks up a different scale factor depending on whether it is in the higher frequency groups G1 or G3, or the lower frequency groups G2 or G4. Specifically, we find
\begin{equation}
    \dot\phi_{\pm i} = \bigg(\frac{\frac{1}{N}\tilde\mu + \tilde\omega + \tilde\sigma\braket{\omega'}_{\pm/0}}{(\tilde\omega + \tilde\sigma\braket{\omega'}_{\pm/0}) (1-\tilde\beta\tilde\mu)}\bigg)(\tilde\omega + \tilde\sigma\omega'_i)\, ,
\end{equation}
where $\dot\phi_{\pm i}$ is the velocity of an oscillator either in one of the faster groups (G1, G3) if $+$ or one of the slower groups (G2, G4) if $-$. Likewise, $\braket{\omega'}_{\pm/0}$ is the appropriate average of $\omega'$, with the $\pm/0$ subscript indicating the way groups should be paired depending on the topology. We define $\braket{\omega'}_+$ to be the average of $\omega'$ taken over the two fast groups G1 and G3. Likewise $\braket{\omega'}_-$ is understood as the average of $\omega'$ over the two slow groups G2 and G4. Finally, $\braket{\omega'}_0$ is understood as the average of $\omega'$ over a fast group and a slow group, which is just zero since the two distributions always add up to a Gaussian with zero mean. That is, we intend $\braket{\omega'}_{\pm/0}$ to mean that, in the high-high topology scenario, the term is $\braket{\omega'}_\pm$, and so takes the same sign as the $\dot\phi_i$ term under consideration, while in the high-low topology scenario the term is $\braket{\omega'}_0 = 0$ regardless of $\dot\phi_{+i}$ or $\dot\phi_{-i}$. The dense notation allows for succinctness in the expression, since the equations are so symmetric.

In the high-low topology scenario, the doubly-synchronized and doubly-unsynchronized solutions both have Gaussian intrinsic frequency distributions, but they have different variances. The variance in the doubly-unsynchronized state is
\begin{equation}
    \sigma^2 = \bigg(\frac{\frac{1}{N}\tilde\mu + \tilde\omega}{\tilde\omega(1-\tilde\beta\tilde\mu)}\bigg)^2\tilde\sigma^2 \, . \label{eq15}
\end{equation}
We can then find the critical coupling using the same equation as before. Importantly, for the high-low topology scenario we are only interested in whether the scale factor (the term in parentheses in Eq.~\ref{eq15}) is greater or less than one. If greater than one, the completely unsynchronized frequency distribution is wider than the completely synchronized distribution and the unsynchronized distribution will then be more difficult to synchronize than it will be to maintain the doubly-synchronized state, which would indicate bistability between those states.

If the scale factor is less than one, inducing synchronization in the doubly-unsynchronized state will be easier than maintaining a fully-synchronized state of both groups. The question then naturally arises as to what happens when synchronization occurs in the doubly-unsynchronized state. In our previous study of this model without the competition dynamics in the bath levels, we observed a similar situation, with the scale factor being $(1-\tilde\beta\tilde\mu)^{-1}$. When the coupling in that model was such that neither the synchronized nor the unsynchronized states were possible to maintain in steady state, when the scale factor was less than one (as it was in the negative feedback case since $\tilde\mu\tilde\beta<0$) the system settled into an oscillatory fluctuation of the order parameter, transiently moving into and out of the synchronized state. In this model, however, there is another state for the system to relax to, which is the state where one group is synchronized and the other is not. 

We also note that the condition under which the given scale factor is less than one is
\begin{equation}
    -N\tilde\beta > \tilde\omega^{-1}\, . \label{ConstraintCompetitionInequality}
\end{equation}
The intuitive reason for this condition is as follows. In the limit where $\tilde\omega$ is very large, all oscillators essentially turn at the same velocity $\tilde\omega$ with very small variations due to $\omega_i'$ and $R_i'$. Thus, the equilibrium distribution of the bath resources will be essentially uniform. The competition dynamics of the model in the bath layer then become entirely negligible, and so we simply have the resource-constrained dynamics studied in our previous article \cite{Kroma-Wiley2021}. In that model, only the sign of $\tilde\beta$ controlled whether the scale factor was greater or less than one, and that is precisely what we find in Eq.~(\ref{ConstraintCompetitionInequality}) when $\tilde\omega \rightarrow \infty$.

The other limit is when $\tilde\beta\rightarrow 0$. Since $\tilde\omega>0$, when $\tilde\beta\rightarrow 0$ we have a scale factor greater than one. In the same $\tilde\beta\rightarrow 0$ limit the resource levels become identical to the bath levels, since there is no resource consumption, and so $R_i$ just equilibrates to $B_i$. Since the bath dynamics prefer to send resources toward faster oscillators, which then induce greater velocities in those oscillators, a positive feedback loop arises. In a positive feedback situation, when a faster-moving oscillator is pulled toward the mean frequency by the coupling to other oscillators, it slows, and therefore resources move away from it, slowing it further. Thus, once the synchronized state arises, it gradually self-reinforces as the resources redistribute. This is precisely a bistable scenario, as expected by the magnitude of the scale factor being greater than one in this limit.
In the high-low topology scenario simulations we presented above, we were always in the regime with scale factor less than one, so synchronization is easier from the doubly-unsynchronized state than desynchronization from the doubly-synchronized state. 

In the doubly-unsynchronized state of the high-high bath topology, things are more complicated. The different scale factors for the high frequency and low frequency oscillators mean that the resultant distribution of intrinsic frequencies will no longer be Gaussian. The high frequency oscillators have a smaller scale factor than the low frequency oscillators due to $\braket{\omega'}$ being greater for the high frequency oscillators than for the low frequency oscillators, and the scale factor decreases monotonically as $\braket{\omega'}$ increases. Thus, the two halves of the initially Gaussian intrinsic frequency distribution will draw near one another in the high-high topology scenario. Further, since we are in the less-than-one scale factor regime, the widths of the natural frequency distributions will also narrow. Thus, synchrony out of the doubly-unsynchronized state will be easier in the high-high topology scenario than in the high-low topology scenario due to its narrower distribution of intrinsic frequencies. 

\subsection{Singly-Synchronized}
The one-group-synchronized state is considerably more difficult to analyze than either of the states we have considered above. For that reason, the bulk of the analysis can be found in the Supplement. 
In the end, we find that when one pair, say \{G1,G2\}, is synchronized, the frequencies of the oscillators in the other group will be
\begin{equation}
    \dot\phi_{\pm i} = \big(1-\tilde\beta\tilde\mu)^{-1}\bigg[\frac{\tilde\mu \braket{B'}_{\pm} + \tilde\omega + \tilde\sigma\braket{\omega'}_\pm}{\tilde\omega + \tilde\sigma\braket{\omega'}_{\pm}}\bigg]\big(\tilde\omega + \tilde\sigma\omega_i'\big)\, ,
\end{equation}
where $\braket{\cdot}_{\pm}$ indicates averaging either over G3 ($+$) or G4 ($-$), remembering that G3 oscillators will tend to be faster on average than G4 oscillators due to the $\omega'$ distributions. Further, $\dot\phi_{\pm i}$ indicates the phase velocity of oscillator $i$ which is a member of G3 if $+$ or G4 if $-$, so this expression for $\dot\phi$ depends on whether we are talking about G3 or G4 oscillators. Finally, $\braket{B'}_{\pm}$ can be found by numerically solving the pair of equations
\begin{align}
    \braket{B'}_{\pm}^2 + \braket{B'}_\pm\bigg[4\frac{\tilde\omega}{\tilde\mu} &+ \frac{2\tilde\sigma\braket{\omega'}_\pm}{\tilde\mu} - \braket{B'}_{\mp}\bigg] \\
    &- \frac{4}{N}\bigg[\frac{\tilde\omega}{\tilde\mu} + \frac{\tilde\sigma\braket{\omega'}_\pm}{\tilde\mu}\bigg] = 0 \nonumber
\end{align}
simultaneously for $\braket{B'}_+$ and $\braket{B'}_-$.

For the oscillators in the synchronized group, their true frequencies are all equal, but we are still interested in the distribution of their intrinsic frequencies, since this will impact when phase transitions occur. Assuming we have already solved the above pair of equations for $\braket{B'}_\pm$, we know that the oscillators in G1 share the remaining resources equally among themselves, and so the only variance in natural frequency comes from the built-in variance in the $\omega'$ terms. The same can be said for the G2 oscillators. However, although the natural frequencies are all drawn from Gaussian distributions with variance $\tilde\sigma^2$, the means of the distributions have shifted by different amounts for the G1 versus the G2 oscillators. In the high-high topology, for example, we know that G1 and G3 share a combined pool of $1/2$ bath resources (since $B'_\text{tot} = 1$). Once we have found $\braket{B'}_+$, the average resource level in G3, we know that G3 has, in total, $N\braket{B'}_+/4$ total resources, and so G1 must contain the remainder, evenly divided among all its oscillators:
\begin{align}
    B'_{i1} &= \bigg(\frac{1}{2} - \frac{N\braket{B'}_+}{4}\bigg)/\frac{N}{4} \nonumber \\
    &= \frac{2}{N} - \braket{B'}_+\, ,
\end{align}
and by the same argument, $B'_{i2} = \frac{2}{N} - \braket{B'}_-$. In the high-low topology scenario, G1 is coupled with G4, rather than G3, and so the subscripts are changed accordingly.
Thus, although the spreads of the distributions are known, their mean values are shifted differently depending on the bath topology.

Now in order to validate our intuitions about the behavior, we will need to make an approximation, as the equations to solve for the average bath levels in the singly-synchronized case are too complicated otherwise. We will assume that the separation between the average velocities of the fast and slow oscillators in the singly-unsynchronized case is small. The separation will be small either when $\tilde\sigma$ is small, meaning that the Gaussian distribution of intrinsic frequencies has little variance, or when $\xi$ is small, meaning an only slight division of fast versus slow oscillators between G1 and G2 (and likewise G3 and G4). The details of the derivation can be found in the Supplement. In the end, we find that the average resource levels are given, approximately, by
\begin{equation}
    \braket{B'}_\pm \approx \frac{1}{N} + \frac{\tilde\sigma\braket{\omega'}_\pm}{N\tilde\omega}\frac{1}{1+\frac{\tilde\mu}{N\tilde\omega}}\, .
\end{equation}

\subsection{Comparison of Frequency Distributions}
Now having all of the equilibrium bath levels in hand, we can begin comparing the intrinsic frequency distributions in each synchronization state in order to understand the behavior of the critical couplings to transition between states.
First, we just compile the $\dot\phi$ equations for all the unsynchronized populations, and the intrinsic frequency distributions (i.e., the expressions for $\dot\phi$ without the coupling term) for all the synchronized populations.

For the doubly-unsynchronized state, both \{G1,G2\} and \{G3,G4\} are unsynchronized, and so have the same frequency distribution. We have
\begin{equation}
    \dot\phi_{\pm i} = \bigg(\frac{\frac{1}{N}\tilde\mu + \tilde\omega + \tilde\sigma\braket{\omega'}_{\pm/0}}{(\tilde\omega + \tilde\sigma\braket{\omega'}_{\pm/0}) (1-\tilde\beta\tilde\mu)}\bigg)(\tilde\omega + \tilde\sigma\omega'_i)\,. \label{Eq:ScaleFactor}
\end{equation}

For the singly-synchronized state, we have a synchronized group and an unsynchronized group. The unsynchronized group will have
\begin{equation}
    \dot\phi_{\pm i} = \frac{\omega + \tilde\sigma\omega_i'}{1-\tilde\beta\tilde\mu}\bigg[\frac{\tilde\mu \braket{B'}_{\pm} + \tilde\omega + \tilde\sigma\braket{\omega'}_\pm}{\tilde\omega + \tilde\sigma\braket{\omega'}_{\pm}}\bigg]\, ,
\end{equation}
where
\begin{equation}
    \braket{B'}_\pm \approx \frac{1}{N} + \frac{\tilde\sigma\braket{\omega'}_\pm}{N\tilde\omega}\frac{1}{1+\frac{\tilde\mu}{N\tilde\omega}}\, . \label{1sBathApprox}
\end{equation}
For the synchronized group, the intrinsic frequency $\nu_i$ will be given by
\begin{equation}
    \nu_{\pm i} = \tilde\omega + \tilde\mu\bigg(\frac{2}{N} - \braket{B'}_{\pm/\mp} \bigg) + \tilde\sigma\omega'_i\, ,
\end{equation}
where $\braket{B'}_{\pm/\mp}$ extends our previous notation used above to now indicate that it depends on the topology being high-high or high-low.

Finally, in the doubly-synchronized state, the variance in the intrinsic frequency of all oscillators (which is all that matters for synchrony) will be $\tilde\sigma^2$.

\subsection{Synchronization Transitions}
Now let us consider the consequences of these equilibrium distributions with respect to our system overview simulation in Figure \ref{SystemOverview}. 
The most prominent feature we observe in Fig.~\ref{SystemOverview} is that, in the high-high topology scenario, it is harder (i.e., requires a greater coupling strength) to transition from the singly-synchronized to the doubly-synchronized state than it is to transition from the doubly-unsynchronized to the singly-synchronized state. Since the critical coupling strength will depend on the velocity distribution in the initial state, not the final state, the relevant velocity distributions are those of the unsynchronized group in the doubly-unsynchronized and singly-synchronized states.

In the doubly-unsynchronized state, we have already discussed that the scale factor in Eq.~\ref{Eq:ScaleFactor} will tend to slide the natural frequency distributions of the high and low frequency populations together toward their mutual midpoint, since the low frequency population has a larger scale factor than the high frequency population. We see sliding of the natural frequency distributions as well in the singly-synchronized state. Thus, the question is how to measure the effect of these shifts on the critical coupling strength. Put another way, we have seen already that when the only change is in the width of the natural frequency distribution, the effect on the critical coupling strength is easy to predict using Kuramoto's original result \cite{Kuramoto1975}. However, when parts of the distribution are shifting left or right, the effect is not only a change in the width of the distribution, but a change in its shape, which is much more complex.

Finding the precise shape of the probability distributions of natural frequencies when parts of the distribution are differently scaled/shifted is generally quite challenging, so we will use a heuristic instead. We will suppose that the ease of synchrony is, at a first pass, controlled by the difference in average velocity between the on-average-faster oscillators in G1 and the on-average-slower oscillators in G2 (or G3 and G4). Thus, we will compare these average differences between the two states, doubly-unsynchronized and singly-synchronized.
In the doubly-unsynchronized case, we find
\begin{equation}
    \Delta\braket{\dot\phi}_{\pm} = \frac{\tilde\sigma\Delta\braket{\omega'}_\pm}{1-\tilde\beta\tilde\mu}\,.
\end{equation}
By comparison, in the singly-synchronized case, we find
\begin{equation}
    \Delta\braket{\dot\phi}_\pm = \frac{\tilde\sigma\Delta\braket{\omega'}_\pm + \tilde\mu\Delta\braket{B'}_\pm}{1-\tilde\beta\tilde\mu}\,. \label{eq26}
\end{equation}
Now, as we can see from Equation (\ref{1sBathApprox}), $\Delta\braket{B'}_\pm$ will be positive, and so the distance between the mean natural frequencies of the fast versus slow oscillators increases in the singly-synchronized state compared to the doubly-unsynchronized state. Thus, we were correct in expecting that synchrony would be harder to obtain in the high-high topology scenario.

In the high-low topology scenario, our intuition was the reverse, that transitioning from the singly-synchronized state to the doubly-synchronized state would be easier than from the doubly-unsynchronized state to the singly-synchronized state. Let us see if our intuition was correct.
In the singly-synchronized state, let \{G1,G2\} be the synchronized pair. Then G1 \& G2 oscillators will turn at the same velocity, and so will have equal biases in the bath layer. Thus, the high-high versus high-low topology does not impact the bath dynamics. Since the singly-synchronized average bath levels are independent of topology, the $\Delta\braket{\dot\phi}_\pm$ term remains unchanged for the singly-synchronized case. However, in the doubly-unsynchronized case, we now have
\begin{equation}
    \Delta\braket{\dot\phi}_\pm = \frac{\tilde\sigma\Delta\braket{\omega'}_\pm}{1-\tilde\beta\tilde\mu}\bigg(1+\frac{\tilde\mu}{N\tilde\omega}\bigg)\,. \label{eq27}
\end{equation}
To see how these compare, we compute $\Delta\braket{B'}_\pm$ using Eq.~\ref{1sBathApprox}, which is
\begin{equation}
    \Delta\braket{B'}_\pm = \frac{\tilde\sigma\Delta\braket{\omega'}_\pm}{N\tilde\omega}\frac{1}{1+\frac{\tilde\mu}{N\tilde\omega}}\,.
\end{equation}
Plugging this into the $\Delta\braket{\dot\phi}_\pm$ equation \ref{eq26}, we find
\begin{equation}
    \Delta\braket{\dot\phi}_\pm = \frac{\tilde\sigma\Delta\braket{\omega'}_\pm}{1-\tilde\beta\tilde\mu} \bigg(1+\frac{\tilde\mu}{N\tilde\omega}\frac{1}{1+\frac{\tilde\mu}{N\tilde\omega}}\bigg)\,. \label{deltaPhi1s}
\end{equation}
Now we notice that the first fraction of Eq.~\ref{deltaPhi1s} is equivalent to that of the doubly-unsynchronized version in the high-low topology, Eq.~\ref{eq27}, while the term in parentheses is clearly always smaller in the singly-synchronized case compared to the doubly-synchronized case. Thus we see again that our intuitions are supported by our mathematical derivations.

We have now accounted for the two most apparent features of our system overview plots in Figure \ref{SystemOverview}, and we have found them to hold regardless of our choices of system parameters, consistent with our intuitions.
Moreover, now that we have these equations in hand, it becomes clearer how we can exacerbate these effects. In the high-high topology scenario, for example, the larger $\Delta\braket{B'}_\pm$ can be made, the greater will be the difference in critical couplings. We see from Eq.~(\ref{1sBathApprox}) that our options for accomplishing this outcome are somewhat limited: we either have to decrease $\tilde\omega$ or increase $\tilde\sigma$ (or increase $\tilde\mu$, but there are diminishing returns under that approach since for large $\tilde\mu$ the $\tilde\mu$ dependence drops out). However, since we require that all $\tilde\omega + \tilde\sigma\omega'_i$ terms be positive, we are fundamentally limited in how great the difference can be made.
By comparison, the difference in the high-low topology scenario can be made arbitrarily great, since the second term of Eq.~(\ref{deltaPhi1s}) can be made arbitrarily small by increasing $\tilde\mu$. 

To complete our discussion, we consider the synchronized populations in the singly-synchronized and doubly-synchronized cases, since these will inform us about the critical coupling strength for the downward transitions. We have already found the bath levels for both cases. The doubly-synchronized case has uniform bath levels for all oscillators, and so the intrinsic frequency distribution will simply be a Gaussian distribution with variance $\tilde\sigma^2$. For the singly-synchronized case, we found that the variance of the intrinsic frequency distribution was the same $\tilde\sigma^2$, but that the means had shifted since the average bath level was no longer equal between the faster and slower oscillator populations. If we again compute the difference in average intrinsic frequency between the faster and slower groups of oscillators, we find
\begin{equation}
    \Delta\braket{\nu}_\pm = -\tilde\mu\Delta\braket{B'}_{\pm/\mp} + \tilde\sigma\Delta\braket{\omega'}_\pm\,.
\end{equation}
We thus see that, in going from the singly-synchronized to the doubly-unsynchronized state, compared to going from the doubly-synchronized to the singly-synchronized state, the difference in mean natural frequencies is smaller in the high-high topology scenario and larger in the high-low topology scenario. There is, however, no definitive ordering relation between these synchronized intrinsic frequency distributions and the unsynchronized velocity distributions, since the unsynchronized distributions generally depend on $\tilde\beta$, while these synchronized distributions do not, and so depending on the value of $\tilde\beta$, the ordering may flip. Thus, looking at the up-left versus down-right hysteretic transitions near the boundaries between phases is a more complicated problem, and is more sensitive to the parameters involved.

\section{Discussion}
In this paper, we have presented a study of a resource-competition synchronization model. Due to the considerable size of the parameter space, we restricted ourselves to analyzing a designed network topology in which we had a strong intuition about the expected behavior of the system. We then analyzed the equilibrium frequency distributions in each of the synchronization states possible in the designed model, thereby validating our intuitions about the system behavior. In addition, the analysis made clear the relevant parameters involved in altering the shapes of the boundaries between phases in $\{K_{12},K_{34}\}$ space. The analysis also made clear that the most important feature consistently involved in generating the boundaries found here is simply the average of the $\omega'_i$ terms within a group and, in particular, the differences of these averages between relevent groups (i.e., groups connected in either the phase coupling matrix $A$ or the bath coupling matrix $C$.) Since $\omega'_i$ in this model is essentially just a parameter which indicates different natural frequencies in the presence of equivalent resource availability, it amounts to a difference in ``firing sensitivity" in a neuronal context. Thus, we expect that in any physical system with resource competition in which populations with fairly different firing sensitivities are coupled together, either by way of phase interactions or by way of a shared resource pool, the observed behavior should be consistent with the qualitative features found in this model.

One limitation of the present work is that it considers strongly-divided community structure within each layer. While many real networks do exhibit strong community structure, so also many do not. Thus, further work on this model should extend its realm of applicability by considering more general network structures, or at least other topologies of biological interest, such as small-world, core-periphery, and real-world-data networks. 

\section{Relevance for Neural Systems}
This model suggests an interesting avenue of future work on the correlation between functional and structural connectivity in the brain. In prior work, it has often been observed that functional connectivity exists between regions of the brain which lack any direct white matter tracts (i.e., structural connections) \cite{Hagmann2008,Honey2009,Skudlarski2008,Liang2013,Liegeois2020}. The reason for such correlations has largely been attributed to the influence of an intermediate region of the brain which connects the two regions indirectly, and so causes a correlation to exist in their behavior without a direct line of communication \cite{Damoiseaux2009}. While such third-party regions are undoubtedly important, this model suggests resource flux throughout the brain as another candidate communication mechanism.

Put another way, functional connectivity is measured using BOLD fMRI, which is not a direct measure of activity, but a measure of oxygen density. Oxygen density, then, is a consequence of resource redistribution, and this redistribution is correlated with activity, giving the indirect measure of interest. However, this presentation seems to suggest a unidirectional means of communication, in which neural activity affects resource distribution, but resource distribution does not significantly affect neural activity. The model presented here simply suggests that perhaps there is a feedback mechanism in which resource distribution does marginally affect neural activity. Further, we show that if such a feedback does exist, then the feedback itself can be responsible for inducing correlations in activity that we would otherwise be inclined to attribute to direct or indirect neuronal interactions. 

This is, of course, not the first time such a correlation has been studied. Prior work has  demonstrated that regional cerebral blood flow (rCBF) is correlated quite strongly with resting functional connectivity strength (rFCS) \cite{Liang2013}. That is, the regions of the brain which show the strongest average functional connectivity to other regions of the brain also tend to be the regions of the brain which have the highest throughput of arterial blood flow. This relation is intuitively consistent with the model presented here because the regions of the brain with the greatest rCBF will, upon modulating vessel size, cause the greatest change in blood flow to the remaining regions of the brain, and therefore will also be very significant nodes in the resource-distribution network of the brain. Thus, nodes with high functional connectivity are highly salient not only in the functional connectivity network, but also in the resource-distribution network, and the model we present here suggests that these two may jointly contribute to high functional connectivity.

In all, it seems there is no reason to reject the feedback mechanism out of hand, and this work suggests that the feedback need not necessarily be especially strong for its consequences to be observed. In the language of the model, the feedback is characterized by the $\mu$ parameter, which relates internal resource supply to firing rate, and many of the major results presented in this paper have linear dependence on $\mu$. This model, then, serves as a baseline dynamical system which can capture the effects of such feedback. Further work, then, may be directed toward the study of this model in network topologies which are more brain-like, and using parameter values which are more biophysically motivated. It is our hope that, when such choices are made, this model will provide some additional explanatory power in the estimation of functional connectivity from structural connectivity and vascular connectivity jointly.

\noindent \textbf{Acknowledgements.}
This work was primarily supported by the Army Research Office (Grant No. W911NF-18-1-0244), the Paul G. Allen Foundation, and the NSF through the University of Pennsylvania Materials Research Science and Engineering Center (MRSEC), Grant No. DMR-1720530. P.J.M. also acknowledges support from the James S. McDonnell Foundation 21st Century Science Initiative - Complex Systems Scholar Award Grant \#No. 220020315. The content is solely the responsibility of the authors and should not be interpreted as representing the official views or policies, either expressed or implied, of any of the funding agencies. The U.S. Government is authorized to reproduce and distribute reprints for Government purposes notwithstanding any copyright notation herein. 

\bibliography{ResourceCompetitionCitations}
\bibliographystyle{apsrev4-2}
\frenchspacing

\end{document}